\documentclass[prl,twocolumn,notitlepage]{revtex4-1}
\usepackage[english]{babel}
\usepackage[utf8]{inputenc}
\usepackage{amsmath,amsfonts,amssymb,bm}
\usepackage{graphicx}
\usepackage{extarrows}
\usepackage{color}
\usepackage{soul}
\usepackage{comment}

\begin{document}
\title{Synchronization in leader-follower switching dynamics}
\author{Jinha Park}
\affiliation{CCSS, CTP and Department of Physics and Astronomy, Seoul National University, Seoul 08826, Korea}
\author{B. Kahng}
\email{bkahng@snu.ac.kr}
\affiliation{CCSS, CTP and Department of Physics and Astronomy, Seoul National University, Seoul 08826, Korea}
\date{\today}

\begin{abstract}
The features of animal population dynamics, for instance, flocking and migration, are often synchronized for survival under large-scale climate change or perceived threats. These coherent phenomena have been explained using synchronization models. However, such models do not take into account asynchronous and adaptive updating of an individual's status at each time. Here, we modify the Kuramoto model slightly by classifying oscillators as leaders or followers, according to their angular velocity at each time, where individuals interact asymmetrically according to their leader/follower status. As the angular velocities of the oscillators are updated, the leader and follower status may also be reassigned. Owing to this adaptive dynamics, oscillators may cooperate by taking turns acting as a leader or follower. This may result in intriguing patterns of synchronization transitions, including hybrid phase transitions, and produce the leader-follower switching pattern observed in bird migration patterns.
\end{abstract}

\maketitle

\section{introduction}

An intricate leader-follower dynamics associated with the rankings of the agents is often observed in diverse areas, including sports, such as team pursuit in speed skating~\cite{team_pursuit} and track cycling, and biological systems, such as bird flocks~\cite{vicsek,pigeon,leadership}, insect swarms~\cite{swarm1,swarm2}, or fish schools~\cite{tuna}. In team pursuits in speed skating, a team of three skaters races with another team with the goal of overtaking the other team, which starts on the opposite side of the rink. Team pursuit is technically demanding; skaters in a team need to follow each other closely and synchronously in a line to minimize the drag. The leader faces the wind, whereas the followers encounter less wind. To reduce the energy expended, the leader leaves the front of the team and rejoins the team at the rear. This switching behavior helps minimize the total energy cost of the team. Otherwise, the leader will be exhausted quickly. Bird flocks also exhibit similar switching dynamics during seasonal long-distance migrations. It was recently found that juvenile Northern Bald Ibis cooperate by taking turns as leaders and precisely matching their flying times in the trailing and leading positions. This strategy of shuffling positions enables the flock to withstand the physical exertion during migration flights.

Attempts have been made to understand the movement of groups of animals such as fish schools, bird flocks, and insect swarms in terms of statistical mechanics using maximum entropy methods~\cite{max_entropy}, agent-based simulations~\cite{herd}, and analytically tractable models~\cite{nonlocal}, including synchronization models. The Vicsek model~\cite{vicsek} provides useful insights into collective swarming behavior in nature, for example, in bird flocks~\cite{pigeon,tuna}. Agents can swarm by adapting the average motion of nearby agents, which may be interpreted as a synchronization phenomenon. Moreover, the presence of an actual leader agent can generally enhance the cohesive movements of the Vicsek agents~\cite{robots,virtual-leader}. The Vicsek model shares some features with the Kuramoto model (KM) in the aspect that, as the interaction strength increases, a phase transition occurs from the incoherent state to the coherent state, i.e., a collective synchronous state. By contrast, whereas the Vicsek model can exhibit rich spatiotemporal patterns in two~\cite{vicsek,swarm2} or three dimensions~\cite{vicsek2}, the KM is more suitable for complex dynamics on a ring. Nevertheless, the two models play complementary roles in the study of collective behavior in natural systems. However, to the best of our knowledge, these synchronization models do not take into account asynchronous updating of an individual's status at each time.  

\section{model}
\begin{figure}[!b]
	\includegraphics[width=1.0\linewidth]{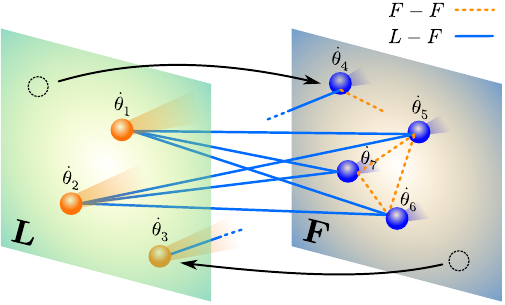}
	\caption{Schematic illustration of the $r$-KM with three leaders and four followers. All oscillators are classified into leaders and followers with fraction $h$ and $1-h$, respectively, according to their angular velocity rankings at each time. Each of followers interact with all the other oscillators, whereas leaders interact with only followers. As a result, the leaders drag the followers, and the followers are dragged by the leaders. Hence, a leader-follower status switching may occur if a follower overtakes a leader, as denoted by a pair of arrows.}
	\label{fig1}
\end{figure}
Inspired by the leader-follower switching dynamics described before, here we modify the original KM with leader-follower switching dynamics. In the model, all oscillators are classified into two groups, the leader group $L$ and follower group $F$ according to their angular velocity rankings at each time step. Each of followers interact with all the other oscillators, whereas each of leaders interact with only followers. According to this rule, circular motions of the leaders are suppressed, in the sense that the leaders, unlike the followers, do not have the advantage of being dragged by others. Thus, this model is called the restricted KM ($r$-KM). We calculate the angular velocity of each oscillator at each instance as follows: 
\begin{align}
\dot{\theta}_i &= \begin{cases} \omega_i + \frac{K}{N} \sum\limits_{j=1}^{N} \sin(\theta_j-\theta_i), &\quad i\in F, \\[5pt]
\omega_i + \frac{K}{\lfloor hN \rfloor} \sum\limits_{j\in F}^{\lfloor hN \rfloor} \sin(\theta_j-\theta_i), &\quad i\in L, \end{cases}
\label{eq:model}
\end{align}
where $\theta_i(t)$ and $\dot \theta_i(t)$ denote the angle and angular velocity of oscillator $i$, respectively. The dot represents the time derivative. The intrinsic frequencies $\{\omega_i\}$ of each oscillators follow the Lorentzian distribution $g(\omega)=\gamma/[\pi(\gamma^2+\omega^2)]$ with $\gamma=0.5$. $K$ is the interaction strength. $N$ is the system size. $\lfloor hN \rfloor$ ($0 \le h \le 1$) is the population of followers. We repeat the update dynamics in \eqref{eq:model} until the system reaches a steady state. As a result of this updating, the leaders drag the followers, and the followers are dragged by the leaders. The detailed rule is presented in Fig.~\ref{fig1}.

It should be noted that switching between leaders and followers occurs. If a follower’s angular velocity exceeds that of the leader, the follower and leader switch groups because the number of elements in each group is fixed. Thus, even though the leader and follower groups are separate, their microscopic compositions are continuously updated by switching oscillators according to their angular velocity rankings. We find that a synchronization transition (ST) from an incoherent state to a coherent state occurs at a transition point $K_c$. Multiple transitions and hybrid STs occur depending on the control parameter $h$. The oscillators with intermediate values of the intrinsic frequency exhibit rapid switching behavior between the two groups. By contrast, those with extreme values of the intrinsic frequency rarely switch unless the coupling constant is sufficiently large beyond the transition point $K_c$. The combined switching intervals span a broad range of values and have a power-law distribution. 

\begin{figure}[!ht]
\includegraphics[width=\linewidth]{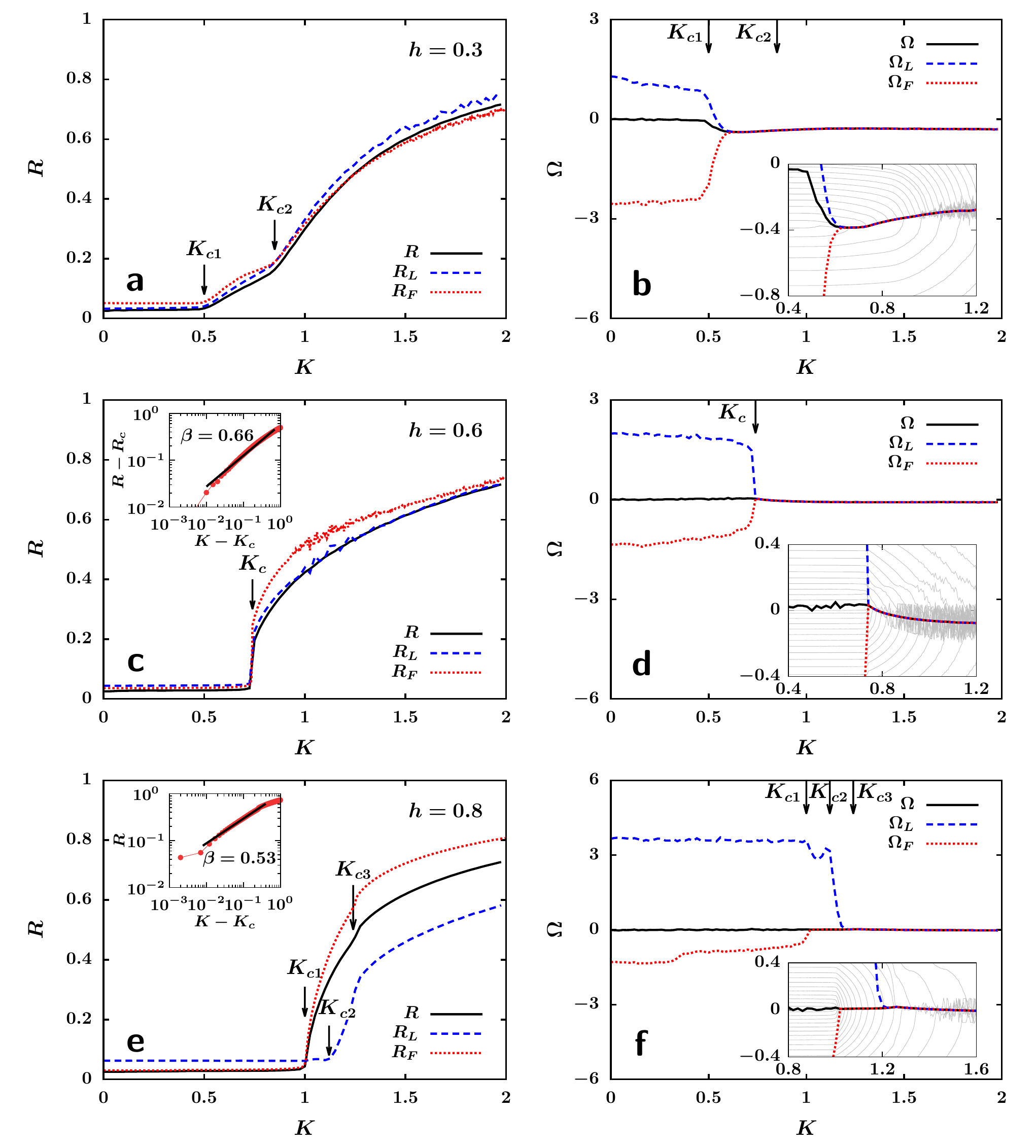}
\caption{Three cases of ST in $r$-KM. Plots of the coherence order parameter $R$ (left column) and traveling wave order parameter $\Omega$ (right column) versus $K$. (a),(b) When the number of followers is small ($h=0.3$), two consecutive transitions occur at $K_{c1}$ and $K_{c2}$, from which $R$ is nonzero, and $\Omega_F$ and $\Omega_L$ converge to zero, respectively. (c),(d) When the number of followers or leaders has an intermediate value ($h=0.6$), the hybrid ST occurs with critical exponent $\beta\approx 0.66$ at $K_c$, from which $R$ is nonzero, and $\Omega$ is zero. (e),(f) When the number of followers is large ($h=0.8$), consecutive transitions occur at $K_{c1}$, $K_{c2}$, and $K_{c3}$, from which $R_F$ increases, $R_L$ increases, and $\Omega$ is zero, respectively. In the insets of (b), (d), and (f), the thin gray curves represent the time-averaged angular velocity $\dot{\bar{\theta}}_i$ of each oscillator $i$ in steady state versus $K$. The curve is drawn for every $20$ oscillators for visualization. Thick black curve represents $\Omega(K)$ for all of the oscillators. Note that at $K_c$ in (d), sudden angular-velocity locking of a macroscopic number of oscillators occurs. By contrast, in (b) and (f), a giant angular velocity cluster grows continuously by merging with other oscillators one by one. For all the panels, $g(\omega)=\gamma/[\pi(\omega^2+\gamma^2)]$, with $\gamma=0.5$, $N=10^3$; the data for a single configuration are used.}
\label{fig2}
\end{figure}

The synchronization transition is characterized by complex order parameters defined as $Z(t)=R e^{i\Omega t+\psi}=\frac{1}{N}\sum_i e^{i\theta_i}$, $Z_F(t)=R_F e^{i\Omega_F t + \psi_{F}}=\frac{1}{\lfloor hN \rfloor}\sum_{i\in F}e^{i\theta_i}$, and $Z_L(t)=R_L e^{i\Omega_L t + \psi_{L}}=\frac{1}{N-\lfloor hN \rfloor}\sum_{i\in L}e^{i\theta_i}$, where $R$, $R_F$, and $R_L$ are the magnitudes of phase coherences of all of the oscillators, followers, and leaders, respectively. In addition, $Z=hZ_F+(1-h)Z_L$. In the incoherent phase, $R=R_F=R_L=0$. In the coherent phase ($R\neq0$), either follower group or both leader and follower groups are synchronized. It is notable that the two groups do not necessarily entrain at a same mean angular velocity in a coherent state. The leader-follower entrainment may follow at a delayed transition point, where the collective phases rotate at a common mean angular velocity $\Omega=\Omega_F=\Omega_L$, but with different phase separations given by $\psi$, $\psi_L$, and $\psi_F$.

\section{Numerical results}
Here, the overall numerical simulation results are presented. The time-averaged coherence $R$ is an order parameter that determines the ST from the incoherent state to the coherent state in Fig.~\ref{fig2}(a), (d), and (g). It is observed that the STs may be categorized into three cases depending on the magnitude of the control parameter $h$: (i) $h < h_\ell$, (ii) $h_\ell < h \le h_u$, and (iii) $h > h_u$. We demonstrate that for cases (i) and (iii), two STs occur consecutively, whereas for case (ii), a hybrid ST occurs. $h_\ell\approx 0.60$ and $h_u\approx 0.72$ are estimated. 

We first consider case (i), specifically $h \ll 0.5$. In this case, all of the followers initially have negative natural frequencies $\omega_i$, whereas the majority of the leaders have positive $\omega_i$. When the coupling constant $K$ is small, switching rarely occurs, and the population arrangement is maintained. Thus, the group angular velocity of the followers, $\Omega_F$, is negative, whereas that of the leaders, $\Omega_L$, is positive; however, $\Omega_L < |\Omega_F|$, as shown in Fig.~\ref{fig2}(b). The two groups rotate in opposite directions. It should be noted that $\lfloor hN \rfloor$ followers interact with all the other oscillators, whereas $N-\lfloor hN \rfloor$ leaders interact only with the followers. Owing to this rule, the synchronization of followers is reinforced but it is weakly disturbed by the leader. Thus, $R_F$ is larger than $R_L$ when $K$ is small. In particular, when $h$ is small, the population of followers is smaller than that of leaders, and the disturbance effect is stronger than when $h$ is large. Thus, the transition point $K_{c1}$ is larger for smaller $h$. The order parameter begins to increase gradually from zero at $K_{c1}$. In the subcritical regime $K < K_{c1}$, the average angular velocities $|\Omega_F|$ and $\Omega_L$ decrease slowly as $K$ increases to $K_{c1}$. $K>K_{c1}$ is a traveling-wave phase. After $K_{c1}$, another transition point exists at $K_{c2}$. Within the interval $[K_{c1}, K_{c2}]$, $\Omega_F$, and $\Omega_L$ are entrained, and they then converge rapidly to a stagnant value of the average angular velocity $\Omega$, as shown in Fig.~\ref{fig2}(b). At $K_{c2}$, a cusp exists in the order parameter $R$ as the system is entrained [Fig.~\ref{fig2}(a)]. As $h$ is increased in (i) but remains below 0.5, the population of followers becomes more similar to that of leaders. Thus, $K_{c1}$ decreases. However, $K_{c2}$ increases because $|\Omega_F|$ becomes more similar to $\Omega_L$. The cusp gradually fades as $h$ increases. When $h$ is close to $0.5$, the follower and leader populations are balanced, and $|\Omega_F|\approx \Omega_L$ in the subcritical regime.

(ii) When $h$ exceeds $0.5$, the population balance breaks down, and $|\Omega_F|$ becomes smaller than $\Omega_L$ in the subcritical regime, in contrast to case (i) [Fig.~\ref{fig2}(e)]. Thus, as $h$ is increased, the disturbance strength becomes larger; i.e., the synchronization is more strongly suppressed, and thus, a larger transition point $K_c$ is needed. As we observed in the restricted Erd\H{o}s-R\'enyi ($r$-ER) model, this $r$-KM exhibits a hybrid ST in interval (ii). The order parameter $R$ behaves as $R-R_0 \sim (K-K_c)^\beta$, where $R_0$ is the jump in the order parameter, and $K_c$ is the transition point~\cite{kcore,Pazo05,metastable,IET}. The exponent $\beta$ is approximately $\beta \approx 0.75$, independent of $h$, which is in contrary to the behavior in the $r$-ER model for hybrid percolation transitions induced by cluster merging dynamics~\cite{rer}. However, $K_c$ depends on $h$: $K_c$ increases as $h$ is increased. The phase and frequency STs occur at the same transition point, $K_c$. $\Omega_F$ and $\Omega_L$ suddenly drop to zero at $K_c$ [Fig.~\ref{fig2}(d)]. These behaviors are associated with the pattern of sudden angular-velocity locking, as shown in the inset. At $K>K_c$, the leader-follower angular velocity may vary depending on the parameter values. It is a traveling-wave phase but has small $\Omega$.

(iii) For $h > h_u$, the order parameter $R$ exhibits multiple transition behaviors [Fig.~\ref{fig2}(e)]. In this regime, because $h$ is sufficiently large, the ST of the entire system is governed mainly by the formation of a synchronized cluster of followers. A continuous ST of the follower group occurs at $K_{c1}$; however, a ST of the leader group follows at a different transition point, $K_{c2}$, where $K_{c2} > K_{c1}$. This is because the leaders do not interact with each other directly, but gather indirectly with the help of a sufficiently large mass of synchronized followers. Although $\Omega_F$ drops to zero at $K_{c1}$, the leader population runs at a different angular velocity, $\Omega_L\neq \Omega_F$, until $K_{c3}$, where $K_{c3}>K_{c1}$ [Fig.~\ref{fig2}(f)]. Hence, $[K_{c1},K_{c2}]$ is a chimera phase, and $[K_{c2},K_{c3}]$ is a standing-wave phase. Finally, the leader group is entrained to the follower group; $\Omega_L$ suddenly drops to zero at $K_{c3}$, and the entire system is frequency-synchronized. The coherent, non-traveling-wave phase lies beyond this point. The order parameter $R$ shows a corresponding increase by a small amount, exhibiting a discontinuous transition. A similar consecutive transition followed by a small jump transition has been observed in the percolation transition in multilayer networks when the interaction between layers is weak~\cite{multiplex}. However, the transition point $K_{c1}$ of regime (iii) weakly varies on $h$, which may be attributed to the finite-size effect of the transition point, because this transition point is determined mainly within the follower group and its population depends on $h$. 

\begin{figure}[!ht]
\includegraphics[width=\linewidth]{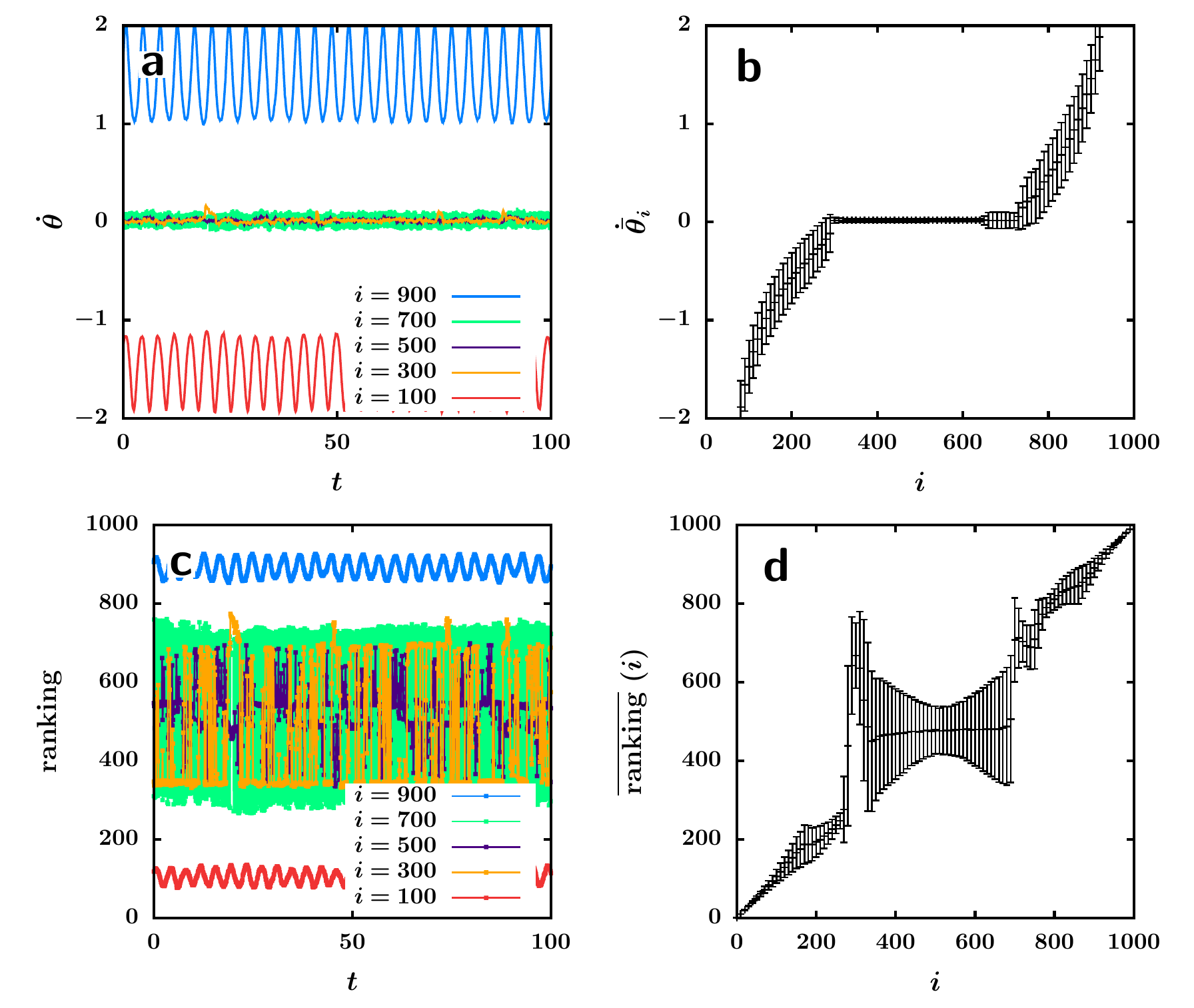}
\caption{Trajectories of (a) angular velocity and (c) ranking as a function of time for given oscillators $i$, where the index $i$ is ordered by increasing natural frequency $\omega_i$. (b) Time average and (d) standard deviation of the quantities in (a) and (c) versus oscillator index $i$. In (a) and (c), a finite fraction of oscillators in the intermediate rankings have angular velocities with small and fluctuating amplitudes. However, the rankings of those oscillators fluctuate greatly. However, the oscillators with small and large indices $i$ show large fluctuations in angular velocity but small fluctuations in rank. Simulations are performed for $N=10^3$. The control parameters are set to $h=0.7$ and $K=1$, which are near the transition point for the hybrid ST. In (b) and (d), the index of $i$ is taken in $10$-oscillator steps.}
\label{fig:outpace}
\end{figure}

\begin{figure}[!h]
\includegraphics[width=\linewidth]{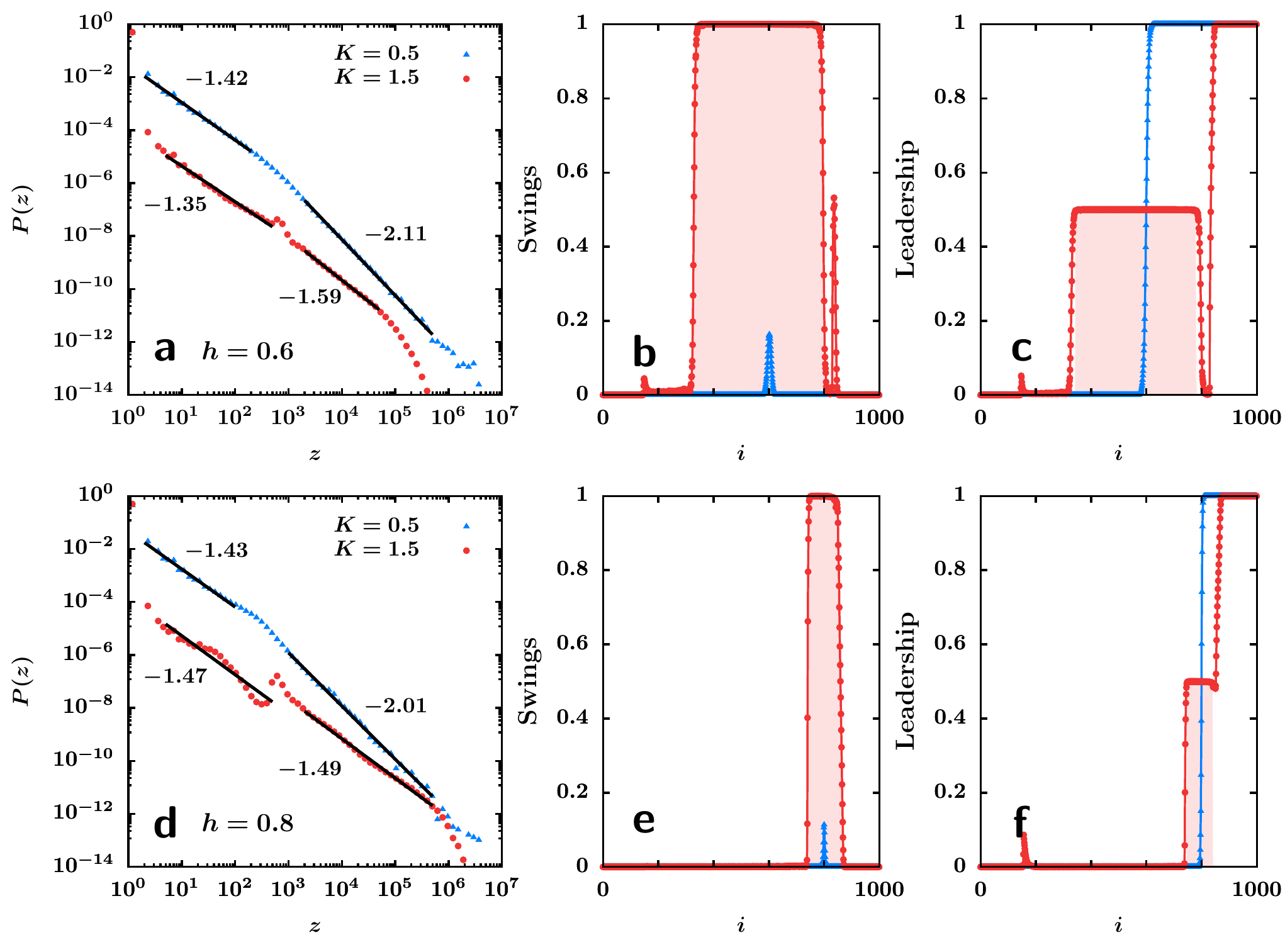}
\caption{Plots of (left column) the interevent time (IET) distribution of leader-follower switches for the $r$-KM, (middle column) each oscillator's frequency of swings, and (right column) participation rate as a leader for a subcritical coupling $K=0.5$ and a supercritical coupling $K=1.5$, in regimes (ii) $h=0.6$ and (iii) $h=0.8$. Note that there are an intermediate swinging class and populations on the sides, which rarely switch. The IET distribution shows crossover behavior between two power-law behaviors. Data in the short IET region are contributed by the thin intervals between the lower and middle and between the middle and upper groups. Data in the long IET region are contributed by the lower group, which also switch, but with small probability. The thin intervals, which build the power laws in the IET distribution, correspond to the kinked regions in the swing and leadership curves. For the subcritical $K$, all the swing curves share a universal soliton-like pattern, and the IET exponents are also independent of $h$. The height of the soliton increases rapidly near the transition point. In the supercritical regime, the swing curve is flat-topped~\cite{nonlocal2} and may be accompanied by additional soliton-like features on the sides. The flat-top width tends to grow with increasing $K$. In regimes (i) and (ii), the swinging population may grow indefinitely until it is nearly equal to the system size, whereas in regime (iii), the maximum width is bounded. In the supercritical regime, the swing curve pattern depends on the detailed parameter values, and supercritical IET exponent values are not universal.}
\label{fig:iet}
\end{figure}

Next, we investigate microscopically the leader-follower switching dynamics. The angular velocity and rank of each oscillator change continuously over time, as shown in Fig.~\ref{fig:outpace}(a) and (c), respectively. This figure shows the angular velocity of a given oscillator $i$ in a system of size $N=1000$ versus time in the steady state. Note that for the oscillators with intermediate $\omega$ ranks, although the changes in the angular velocity, $\dot \theta_i$, are small in (a), and the fluctuations are also small in (b), the changes in their rankings are quite large in (c), and the fluctuations are also large in (d), particularly at both boundaries of the middle group. Moreover, in Fig.~\ref{fig:iet}(b) and (e), the oscillators with intermediate $\omega$ ranks in the shaded region switch their status consecutively with a probability close to unity, spending half of the total time as leaders and the other half as followers. Their population increases as $K$ is increased in the supercritical regime. This result seems to resemble the pattern of bird migration, where all the birds act as a leader at roughly the same rate~\cite{leadership,bode}. By contrast, oscillators on the low or high side of the $\omega$ rankings rarely switch their leader-follower status unless the coupling constant $K$ is large enough that the middle class takes over a wide range of population. Between the unchanging high or low class and the switching middle class, there are thin intervals of oscillators that switch frequently over the follower-leader separation with probability $q$. Let us denote the switch probability distribution as $P(q)$ and assume that it follows a power law, $P(q)\sim q^{-x}$. If we assume that the oscillator motions are uncorrelated, the interevent time (IET) distribution is calculated as $P(z) = \int_0^1 dq P(q) P(z|q) = \int_0^1 dq q^{-x} q(1-q)^{z-1}=B(2-x,z)\sim z^{-(2-x)}$, where $P(z|q)$ is the conditional probability, and $B(2-x,z)$ is the beta function. The case $x=0$ corresponds to the uniform probability distribution, where the exponent of $P(z)$ is two. A nonuniform distribution or the effects of the correlations may change the precise value of the exponent. Consequently, the IETs between two consecutive switching events are heterogeneous, and their distribution has a power-law tail, $P(z)\sim z^{-\alpha}$ [Fig.~\ref{fig:iet}(a) and (d)]. This heavy tail in the IET distribution arises from the dichotomous dynamics and global sorting process. The IET distribution exhibits crossover behavior between short and long IETs. Short IETs are due to the oscillators in the thin intervals between the lower and middle groups and between the upper and middle groups. Long IETs are due to the oscillators in the lower group which switch with small probability. Oscillators in the upper group, on the other hand, do not switch. We find empirically that the exponent of the IET distribution depends on the shape of the frequency distribution $g(\omega)$, and also on the model parameters $K$ and $h$.

\section{discussion}

For the ordinary KM ($h\to1$), the phase transition type is determined by the shape of the frequency distribution $g(\omega)$. A unimodal or bimodal $g(\omega)$ yields a continuous or discontinuous transition, respectively~\cite{kuramoto}. A hybrid ST appears in some special marginal cases, such as
flat $g(\omega)$ classes~\cite{Pazo05,basnarkov,metastable} and a degree-frequency-correlated model on a scale-free network with the degree exponent $\lambda=3$~\cite{dorogovtsev-dfckm}. In these cases, the maximum competition arises among the oscillators and suppresses the formation of a synchronization seed cluster. Consequently, a macroscopic number of oscillators are suddenly entrained at a transition point. In contrast to the previous models, the $r$-KM can produce a hybrid ST if the parameter $h$ is tuned to an appropriate value, even for a unimodal $g(\omega)$. The reason is the asymmetric interactions between the leader and follower groups and the tug-of-war critical switching dynamics. Moreover, the $r$-KM exhibits various other ST patterns, e.g., multiple consecutive transitions. The reason is that the leader and follower groups may be regarded as double layers or communities, where multiple transitions occur naturally when the interaction between the two layers is asymmetric~\cite{multiplex,moreno}. By contrast, traveling waves are absent in the ordinary KM with symmetric $g(\omega)$. In the $r$-KM, however, traveling waves are possible owing to the asymmetric interactions between leader and follower groups, which also appear in the KM with asymmetrically competing interactions~\cite{metastable}. It is notable that the $r$-KM takes into account the temporal adaptations of individual oscillators by allowing oscillators to switch their leader/follower status. Hence, the model produces novel self-organized phenomena, where we find a rapidly switching middle class that is half followers and half leaders, as shown in Fig.~\ref{fig:iet}. This type of temporal organization equalizes the distribution of the duration times in the follower and leader positions; this behavior is similar to that observed in the flight patterns of migrating flocks of birds~\cite{leadership}. Finally, we remark that Eq.~\eqref{eq:model} is invariant under $(\theta,\omega,g) \leftrightarrow (-\theta,-\omega,1-g)$. Hence, an inverse model of the $r$-KM, with a limitation applied to the followers instead of the leaders, yields an identical result except that $g\to 1-g$. In addition, the model has rotational symmetry, and the order parameter curve $R(K)$ is not affected by the translation of the frequency distribution $g(\omega) \to g(\omega-\Omega_0)$, where the distribution is centered at a nonzero mean, $\Omega_0$.

In conclusion, we proposed an oscillator model (the $r$-KM) that includes leader-follower switching dynamics in addition to synchronization. This rank-based status reassignment, which mimics adaptive behavior in complex systems, is unprecedented in synchronization models. This new paradigmatic model exhibits rich dynamical behavior that encompasses a hybrid ST, a traveling-wave phase, standing waves, chimera states, and a power-law IET distribution. These diverse patterns of STs arise from the rich behavior of the model, which includes asymmetric interactions, a multilayer leader-follower perspective, and temporal shuffling. Here, we studied the case of Lorentzian distribution $g(\omega)$, but the results can be extended to other distributions as well.

\noindent\textbf{Methods}
For numerical integrations we applied the Heun's method with discrete time step size of $\Delta t=0.01$. Natural frequencies $\omega_i$ were regularly sampled from the Lorentzian frequency distribution $g(\omega)=\gamma/[\pi(\gamma^2+\omega^2)]$ so that $G(\omega_{i})$ are equally spaced, where $G(\omega)=\int^\omega g(\omega')d\omega'$. The width of the distribution was fixed to $\gamma=0.5$, which gives a continuous transition at $K_c=1$ for the ordinary KM (which corresponds to $h=1$ of the $r$-KM). The traveling wave order parameters $\Omega$, $\Omega_L$, and $\Omega_F$ were obtained by measuring the time averaged rate of change in the complex phases $\psi$, $\psi_L$, and $\psi_F$. Those collective phases were sampled every two consecutive time steps to get rid of the irrelevant high frequency noises. Order parameters were time averaged for a sufficiently long time after reaching a steady state.

\noindent\textbf{Acknowledgements}
This research was supported by the National Research Foundation of Korea (NRF), Grant No.~NRF-2014R1A3A2069005 (BK).

\noindent\textbf{Author contributions}
JP and BK designed the research, conceived the study, carried out the analysis and prepared the manuscript.

\noindent\textbf{Competing interests}
The authors declare no competing interests.

\noindent\textbf{Additional information}
Supplementary Information (SI) includes the detailed self-consistency analysis of the $r$-KM and compares hybrid critical exponents $\beta$ obtained from the simulations and the self-consistency theory.

\bibliographystyle{apsrev4-1} 

\end{document}